# Observations of red–giant variable stars by Aboriginal Australians

Duane W. Hamacher

Monash Indigenous Studies Centre, Monash University, Clayton, VIC, 3800 Australia
Astrophysics Group, University of Southern Queensland, Toowoomba, QLD, 4350, Australia

Email: duane.hamacher@monash.edu

**Abstract**

Aboriginal Australians carefully observe the properties and positions of stars, including both overt and subtle changes in their brightness, for subsistence and social application. These observations are encoded in oral tradition. I examine two Aboriginal oral traditions from South Australia that describe the periodic changing brightness in three pulsating, red–giant variable stars: Betelgeuse (Alpha Orionis), Aldebaran (Alpha Tauri), and Antares (Alpha Scorpii). The Australian Aboriginal accounts stand as the only known descriptions of pulsating variable stars in any Indigenous oral tradition in the world. Researchers examining these oral traditions over the last century, including anthropologists and astronomers, missed the description of these stars as being variable in nature as the ethnographic record contained several misidentifications of stars and celestial objects. Arguably, ethnographers working on Indigenous Knowledge Systems should have academic training in both the natural and social sciences.

## Introduction

Aristotle (350 BCE) wrote that the stars are unchanging and invariable, a position held in academic discourse for two millennia. Aside from the appearance of the occasional "guest star" (novae and supernovae) it was not until the year 1596 that this position was challenged. After nearly two thousand years, careful observations of the star Mira (Omicron Ceti) by David Fabricius in 1596 revealed that its brightness changes over time (Hoffleit 1996). The amplitude and periodicity of these changes were calculated by Johannes Hevelius in 1662, definitively overturning Aristotle's claim and ushering in a new era of variable star research. This is regarded as the established discovery of variable stars by historians of astronomy.

Did Indigenous cultures note these changes and record them in written texts or oral traditions? Researchers have attempted to identify (non–eruptive[1]) variable stars in the texts of ancient sky–watching civilisations, such as pre–Classical Greece (Wilk 1996) and Egypt (Jetsu *et al*. 2013). The evidence from pre–Classical Greece is open to interpretation, but a careful analysis of the Egyptian Cairo Calendar, which dates to the period 1271–1163 BCE, shows that the ancient Egyptians noted the variability of the eclipsing binary star Algol (Omicron Persei) (Jetsu and Porceddu 2015). To date, nothing has been published shat shows any clear evidence that Indigenous peoples observed and recorded stellar variability in their oral traditions.

Astronomical knowledge is a significant component of the culture and cosmology of many Indigenous peoples, particularly Aboriginal and Torres Strait Islander cultures of Australia, with direct importance to anthropological studies (Clarke 2007, Johnson 1998). Given the scholarship showing that many Indigenous peoples were (and are) keen observers of the night sky (Norris 2016), is it possible that oral, sky–watching cultures observed the variability of stars and incorporated this phenomenon into their knowledge systems? To answer this question, I examine oral traditions of Aboriginal Australians.

Aboriginal people have lived in Australia for more than 65,000 years (Clarkson *et al*. 2017), speaking over 350 distinct languages (McConnell and Thieberger 2001), providing a diverse range of deep–time traditions that we can explore for this evidence. Since Aboriginal cultures are oral rather than literate, the laws, social rules, and knowledge are committed to memory and transmitted to subsequent generations through oral tradition (Kelly 2015). Oral traditions consist of cultural narratives that describe the creation of the world by ancestors who established the traditional law that guides the people through their daily lives (Clunies–Ross 1986). Rather than being static in nature, oral traditions and Knowledge Systems are dynamic, incorporating new knowledge and experiences as the people and their environment change over time (Battiste and Henderson 2000).

Multifaceted and multi–layered Indigenous Knowledge Systems contain *a priori* and *a posteriori* forms of knowledge. These explain the origin of natural features, the dynamics of natural processes, and various natural phenomena through deduction, observation, experimentation, and experience. This has application to ecology (Prober *et al*. 2011), astronomy (Cairns and Harney 2003), meteorology (Green *et al.* 2010), geological events (Hamacher and Norris 2009), and physical geography (Nunn and Reid 2016). An anthropological understanding of this knowledge can help us gain a better understanding of the development of Aboriginal cultures, how Aboriginal people understand their worldview, and the myriad ways in which Western academia can learn from these traditional Knowledge Systems for mutual benefit.

Aboriginal and Torres Strait Islander people observe the positions and properties of stars to inform navigation, calendar development, and plant/animal behaviour (Clarke 2014, Hamacher *et al*. 2017). These properties include stellar brightness, colour, relative position with respect to other celestial objects, and position with respect to the horizon. Changes in these characteristics are observed and interpreted to predict weather and seasonal change (Parker and Lang 1905: 73–74). Transient phenomena, including meteors, cosmic impacts, and eclipses, are often incorporated into oral tradition, serving as mnemonics for obeying traditional law and avoiding social taboos (Hamacher and Norris 2010, Hamacher and Goldsmith 2013, Hamacher and Norris 2011a, respectively).

Indigenous and Western systems of knowledge are dynamic and experimental in nature, allowing them to evolve with the introduction of new knowledge (Flavier *et al*., 1995). Traditionally, Indigenous Knowledge has been viewed as inferior to Western science, with centuries of colonisation, subjugation, and oppression reducing or eliminating their influence or presence (Laws 1994). Frameworks for working at the intersection of Indigenous Knowledge and Western Science argue that productive engagement involves breaking down this barrier and moving beyond the comparative focus of pitting Indigenous and Western ways of knowing against each other, or focusing on distinctions between them (Agrawal 1995, Nakata 2010).

This paper is generally positivist in its approach, arguing that Aboriginal oral traditions contain information that was gained through careful, long–term observations of stars. I show that subtle changes in stellar properties (such as variable brightness) were observed by Aboriginal peoples and incorporated into their oral traditions. I accomplish this by analysing two oral traditions from South Australia that were published in the anthropological literature. I show that these traditions describe the variability of the pulsating red–giant stars Betelgeuse (Alpha Orionis), Aldebaran (Alpha Tauri), and Antares (Alpha Scorpii), and allude to the relative periodicities in the stars' variability. The evidence shows that variable stars serve as a mnemonic, reflecting cultural practices and traditional laws, and that these views may have some physiological and psychological basis in human perception. This supersedes the accepted consensus by historians of astronomy that the variability of the stars Betelgeuse, Antares, and Aldebaran was first 'discovered' by Western scientists in the nineteenth and twentieth centuries.

**Oral Tradition 1: Nyeeruna**

Aboriginal oral traditions associated with the Western constellation of Orion and the Pleiades star cluster are ubiquitous across Australia (Johnson 2000), with the notable exception of Tasmania (Johnson 2011). These traditions commonly involve the stars of Orion, who are typically a male hunter or group of hunters, pursuing the Pleiades star cluster, which are commonly associated with a group of women, usually sisters. In some traditions, the Hyades star cluster, which lies between Orion and the Pleiades, serves as a barrier between the man/men of Orion and the women of the Pleiades (Haynes 2000).

The dynamic between these stars is often reflected as Orion pursuing the women of the Pleiades to make them his wives (White 1975). In many traditions of this nature, the women of the Pleiades do not reciprocate this love interest (and are constantly running away from the advances of Orion (Fredrick 2008). The diurnal motion of the stars as they rise in the East and set in the West represents a continual celestial chase. This shares close similarities with Greek traditions of these stars, where Orion the hunter pursues the Seven Sisters of the Pleiades to make them his wives, but is challenged by Taurus the Bull, represented by the Hyades cluster (see Figures 1 and 2 for maps of both constellations).

In the Great Victoria Desert of South Australia, Kokatha communities share an oral tradition that describes the relationship between anthropomorphic ancestors that are represented by Orion, the Hyades, and the Pleiades. The tradition was published by amateur anthropologist Daisy Bates (1921), who spent nineteen years living with Aboriginal communities near Ooldea, South Australia at the start of the twentieth century and re–analysed by Leaman and Hamacher (2014).

In the oral tradition, a man named Nyeeruna is a skilled hunter and a vain womaniser who lives in the sky. He comprises the stars of Orion, in the same orientation as his Greek counterpart (meaning he is upside–down as seen from Australia). He pursues the Yugarilya sisters of the Pleiades across the sky each night in an attempt to make them his wives, a pursuit indicated by the relative diurnal motion of the two star groups. Nyeeruna is prevented from reaching the Yugarilya sisters by Kambugudha, their eldest sister who is represented by the Hyades star cluster.

Kambugudha is protective of her younger sisters and is contemptuous of Nyeeruna. She stands before Nyeeruna, mocking and taunting him while blocking him from reaching the sisters. Nyeeruna is filled with lust and is angry he is being prevented form reaching the sisters. The club in his right hand (Betelgeuse) fills with 'fire magic', ready to throw at Kambugudha. She defensively lifts her left foot (Aldebaran), which also fills with fire magic. She kicks dust into Nyeeruna's face, humiliating him. This causes the fire magic of Nyeeruna's hand to dissipate. Kambugudha then places a row of dingo pups in front of Nyeeruna to shield her and her sisters from his unwanted advances. The pups are represented by the curve of stars consisting of $\pi^{1,2,3,4,5}$, $o^2$, 6, 11, and 15 Orionis, which comprise the stars of Orion's shield in Greek traditions (Figure 3).

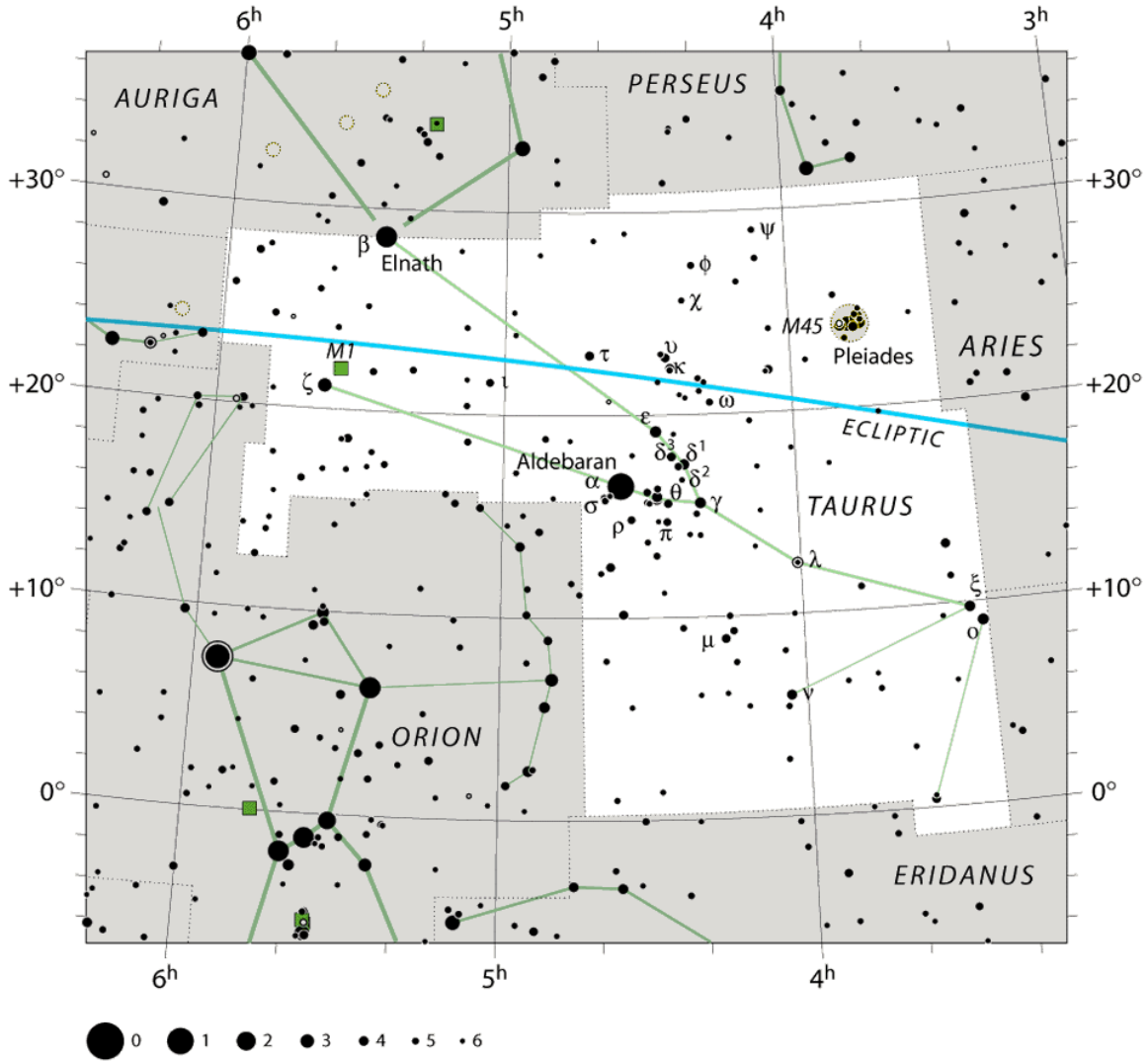

*Figure 1*: Star chart of the constellation Taurus, with Aldebaran, and the Pleiades labelled and Orion to the lower left. Star magnitudes are represented by size, and the x– and y–coordinates are in right ascension and declination. This constellation is seen upside–down in Australia. International Astronomical Union.

Over time, Nyeeruna's magic returns and his club–hand (Betelgeuse) again increases in brightness and 'fire lust' as he pursues the sisters. Kambugudha calls out to Babba, the father dingo, who attacks Nyeeruna while she points and laughs. The timid Yugarilya sisters are frightened and hide their heads until Babba releases Nyeeruna. The surrounding stars join Kambugudha in mocking and laughing at Nyeeruna, who again loses the fire lust of his hand as he faces shame and humiliation. Babba was described by Bates as "the horn of the bull", without specifying if he was Beta or Zeta Tauri. Beta Tauri is brighter at magnitude 1.65, but Zeta Tauri (magnitude 3.01) is closer to Orion and is also variable in nature. It is an eclipsing binary with a variation of 0.1 magnitudes (Harmanec *et al.* 1980). There is no clear indication in the story that Babba is variable, so we cannot make any judgments about this property in the oral tradition.

Variants of this tradition are found across the Great Victoria Desert (Anonymous 1922) and central Australia. A similar tradition to the north of Ooldea involves a man named Njuru as Orion pursuing the sisters of the Pleiades (see White 1975), and several similar variants are found across the Central Desert, including Pitjantjatjara and Yankunytjatjara traditions about Nirunja (Orion) and Kunkarangkalpa (Pleiades), and traditions to the north at Glen Helen Gap (Mountford 1976).

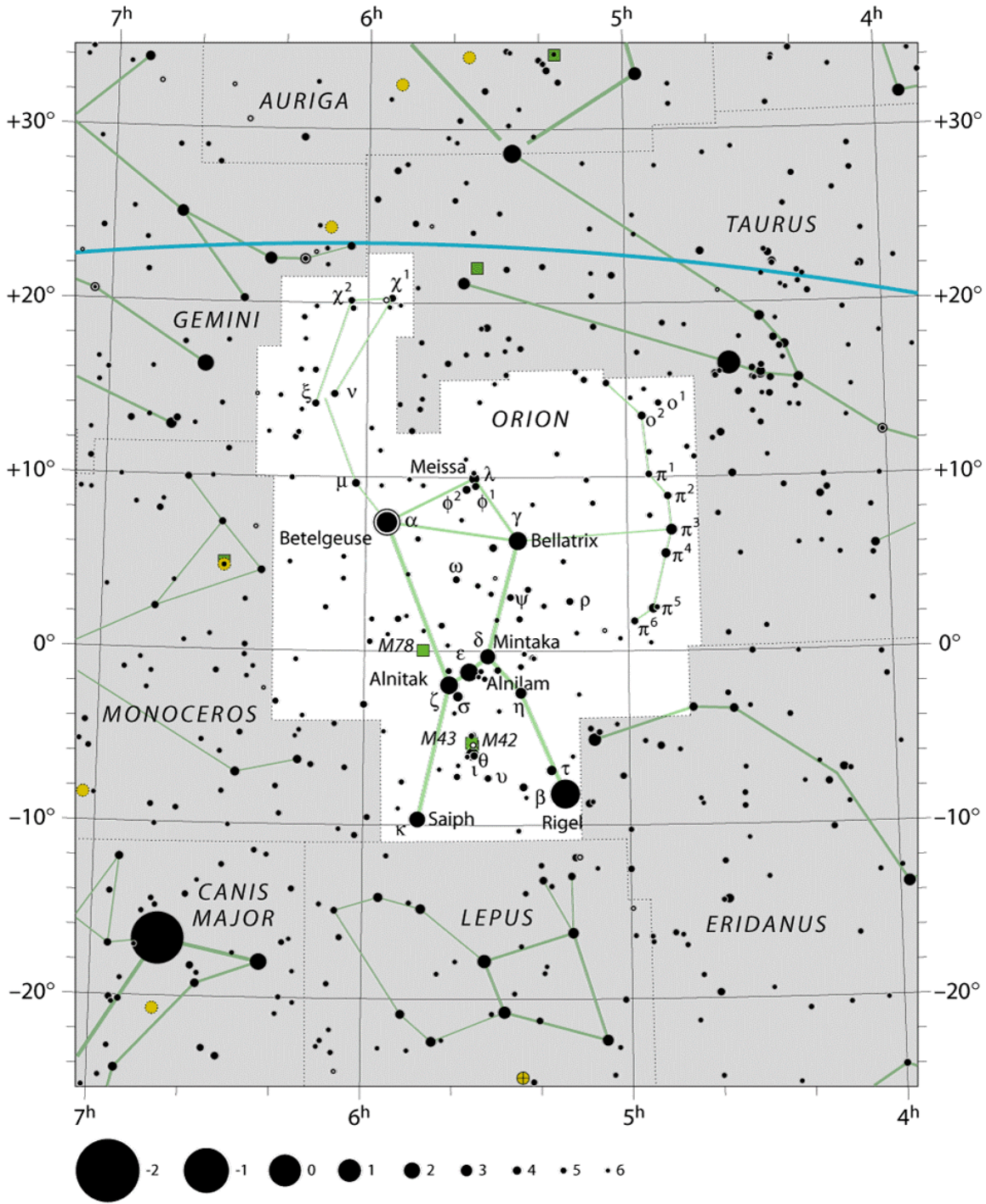

*Figure 2:* Star chart of the constellation Orion, with Betelgeuse labelled and Taurus to the upper right. Star magnitudes are represented by size, and the x– and y–coordinates are in right ascension and declination. This constellation is seen upside–down in Australia. International Astronomical Union.

**Oral Tradition 2: Waiyungari**

In the Lower Murray region of South Australia, young Ngarrindjeri men went through a complex initiation as part of their transition to adulthood (Berndt and Berndt 1993). During the process, male novices (*Narambi*) spend a period of time covered in red ochre and living in isolation, going without food, sleep, or clothes and abstaining from any contact with women. Sexual contact during this period is strictly forbidden. Breaking of this sacred taboo can result in severe punishment for the offending men and women, as well as their families. An element of forbidden love sometimes enhances the Narambi's sexual attractiveness to young women and the narrative serves as a warning to the people about obeying traditional law. The oral tradition describes a young man named Waiyungari (meaning 'Red Man'), which was first recorded by Meyer (1846) and later described by Taplin (1879: 57) and several ethnologists over the next hundred years. A number of variants of the tradition exist across the region (see Clarke 1999), but they adhere to the same primary theme.

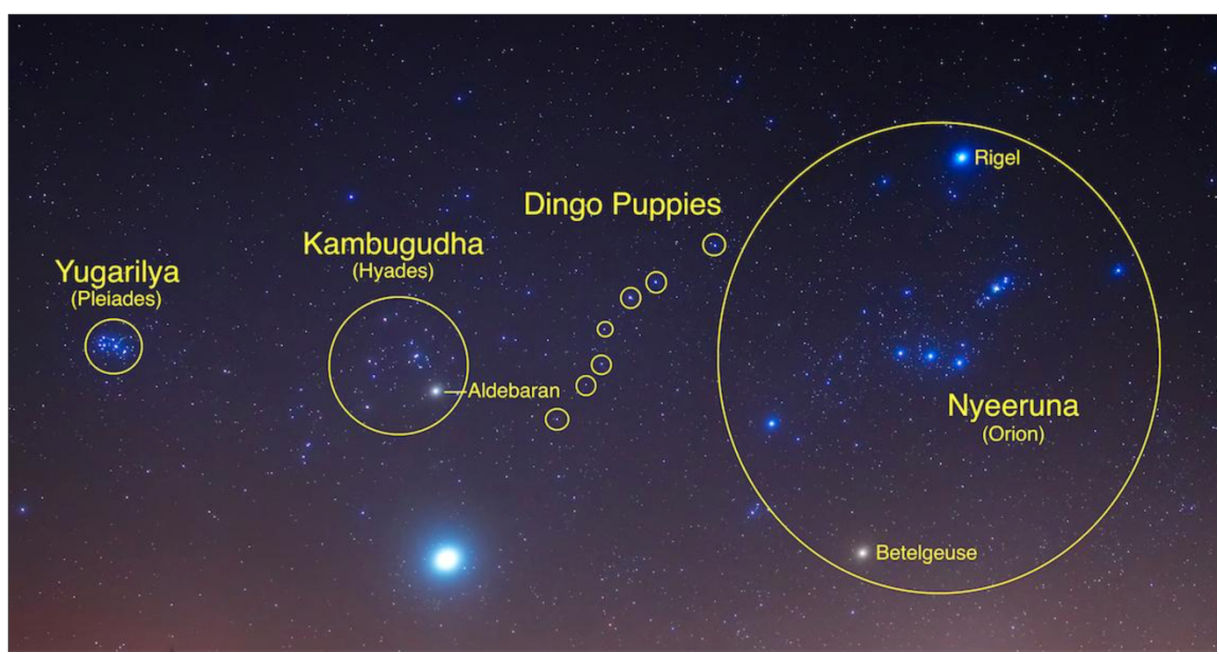

*Figure 3:* Nyeeruna, Kambugudha, and the Yugarilya sisters in the sky. Image reproduced with permission from Leaman and Hamacher (2014).

The narrative describes Waiyungari as a Narambi, covered in red ochre (Clarke 1999). He caught the attention of two women, who were the wives of his brother, Nepeli. The women were deeply attracted to Waiyungari and followed him as he went about his activities in isolation. They approached Waiyungari's hut, took the form of emus and made sounds to draw him out. When he ran out of his hut to pursue the emus, the women morphed into human form and seduced him. Nepeli discovered this betrayal and attempted to exact his revenge by setting ablaze the hut in which his wives and brother slept. The trio managed to escape and ran along the river. To avoid punishment, Waiyungari cast a spear into the Milky Way and pulled himself and the women up into the sky. Waiyungari became a bright red star, signifying the colour of his ochred body. The two women became fainter stars that flank him on either side. They all sit in a canoe in the Milky Way, with the celestial emu to the west. On occasion, Waiyungari brightens and gets 'hotter', increasing the sexual desire of the people. It is during this time the Narambi must refrain from lascivious activities.

The oral tradition was studied by anthropologist Norman Tindale (1935). During his ethnographic fieldwork, Tindale identified Waiyungari as the planet Mars and stated that several "native sources" confirmed this. He was unable to identify the stars that represent the two women. In May 1935, Tindale consulted G.F. Dodwell, the Government Astronomer, who suggested that if Waiyungari was Mars, then the women were probably Jupiter and Venus (Tindale n.d., Tindale 1983: 368). Dodwell's justification was based on symbolism: he argued that the planets wander the sky and occasionally "come into conjunction with Mars, travel with it, and are together overwhelmed by the fiery orb of the sun, re–appearing after a lapse of time as evening stars." Tindale (1983: 369) later discussed the oral tradition with the archaeoastronomer Von Del Chamberlain, who suggested that one of the wives was probably Saturn rather than Venus, as the behaviour of Saturn more closely fits the description provided by Dodwell. In either case, changes in Waiyungari's brightness were attributed to the variable distance of Mars from Earth, which makes Mars' brightness range from magnitude −2.91 to +1.84.

In the 1930s and 40s, husband–and–wife anthropologists Ronald and Catherine Berndt worked with those local Aboriginal communities. The Berndts repeated Tindale's identification of Waiyungari as Mars. They also state that Waiyungari dominates the evening skies in September (Berndt and Berndt 1993). The Berndts describe Waiyungari in terms of symbolic anthropology: he is considered the personification of sexual prowess and fertility, acting as a seasonal marker for the arrival of Spring. His presence high in the sky influences the sexual desire of people and animals. The Berndts (1951: 223) wrote that when Waiyungari is at its brightest, the sexual feelings of men and women are significantly enhanced.

The problem lies with the identification of Waiyungari as Mars. As a planet (meaning "wandering star" in Greek), Mars is not a suitable annual seasonal indicator, as it will not always be in the same position in the sky at a given time of the year. The description of a bright red star in the Milky Way, sitting in the celestial canoe with the emu to the west, appearing high in the evening sky in September clearly points to the red–giant Antares (Alpha Scorpii). Antares is flanked on either side by Tau and Sigma Scorpii, two fainter white stars of comparable brightness (Figure 4). Antares lies near zenith at dusk in September and the trio of stars comprise part of the celestial canoe in the Milky Way. The emu is made–up not of bright stars, but of dark absorption nebulae in the Milky Way between the Coalsack nebula (the head) in the constellation Crux and the galactic bulge (the body), to the West of the canoe (Hamacher 2012; Figure 5).

The identity of Waiyungari as Antares and the two women as Tau and Sigma Scorpii is supported by comparative analysis of the oral traditions of other Aboriginal groups in the region. The adjacent Wergaia people of north–western Victoria describe Antares as Djuit with his two wives represented by Tau and Sigma Scorpii (Stanbridge 1861). In the Arrernte traditions of Central Australia, Antares is a woman covered in red ochre, accompanied by two other women (Tau and Sigma Scorpii), who are trying to avoid the advances of a group of men (Maegraith 1932). In Wurunjerri traditions from central Victoria, Nurong is the brother of the primary creation ancestor, Bunjil. Nurong is represented by Antares and his wives are represented by Tau and Sigma Scorpii (Howitt 1904: 128). The Waiyungari tradition closely resembles an oral tradition from the Clarence River in north–eastern New South Wales, in which a man named Karambal stole another man's wife. In fear of retribution, he hid in a tree. The angry husband found Karambal hiding and set the tree ablaze. Karambal ascended into the sky as smoke to become the star Aldebaran (Clarke 2015). Finally, a Pitjantjatjara tradition from the Central Desert near the Tomkinson Ranges tells of an initiate who was seduced by a young woman. Because of his resent circumcision, he was swollen and they were unable to separate during copulation. Fearing punishment by death for breaking traditional law, they travelled to the sky where they and their tracks became the close–visual binary stars Mu$^1$ and Mu$^2$ Scorpii in the tail of Scorpius (Mountford 1976: 456–459: see Figure 4).

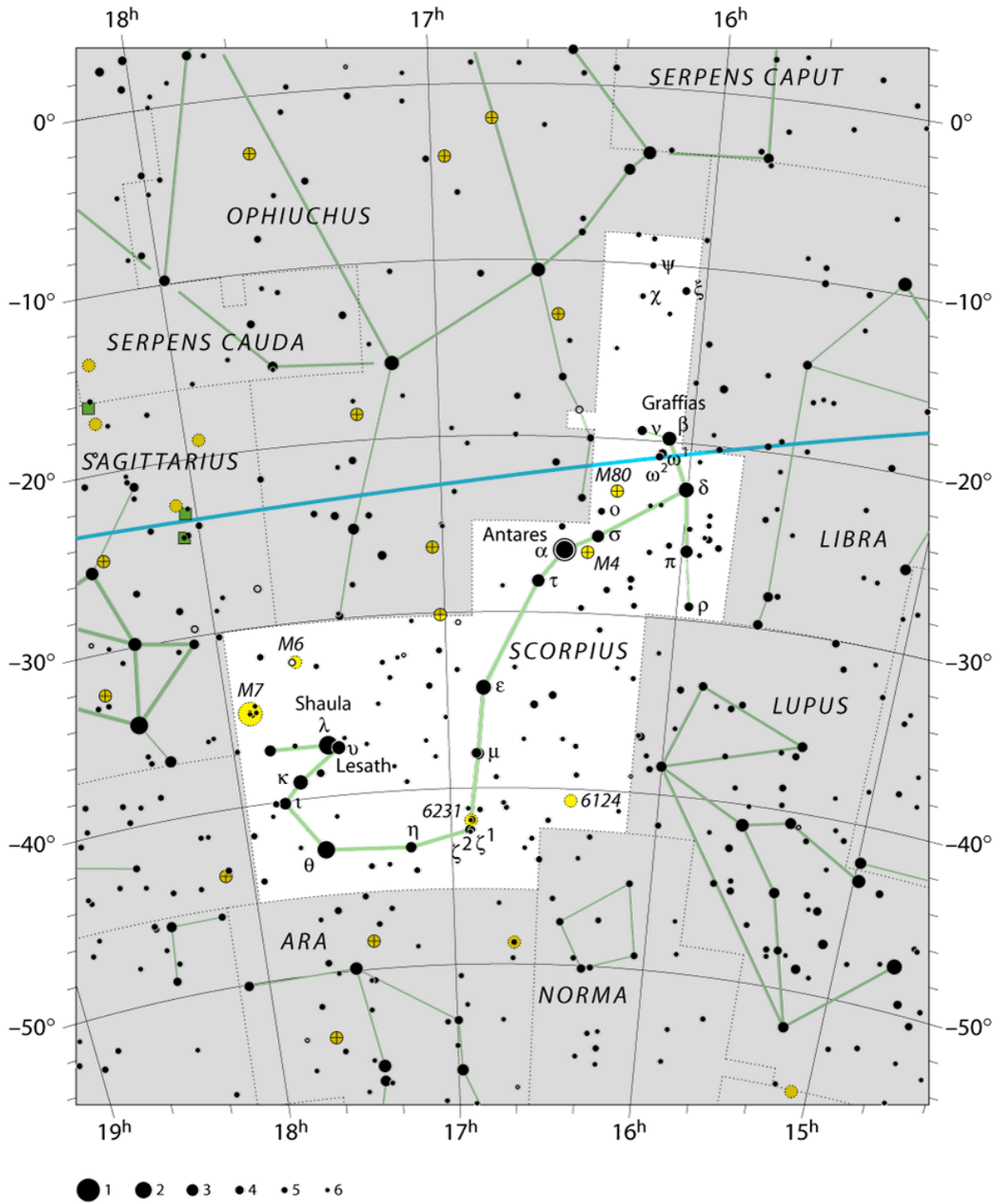

*Figure 4:* *Star chart of the constellation Scorpius, with Antares, Tau Scorpii, and Sigma Scorpii labelled. Star magnitudes are represented by size, and the x– and y–coordinates are in right ascension and declination. This constellation is seen upside–down in Australia. International Astronomical Union.*

Tindale said Waiyungari was Mars and that several Aboriginal people confirmed this. Did the the Aboriginal people physically point out Mars in the sky or did they simply discuss it with Tindale? We do not know exactly when Tindale conducted his fieldwork. Tindale sought the opinion of Dodwell in May 1935. Using the Stellarium astronomy software package [2], we see that Mars was prominent in the sky, to the north of Scorpius, at this time. By September 1935, Mars and Antares reached their closest approach, being separated by only three degrees. Two years earlier in October 1933, Mars was again at its closest approach to Antares. Could this have anything to do with the confusion or misidentification?

Mars and Antares are commonly linked in both Classical astronomy and in the astronomical traditions of Aboriginal groups across Australia. Ares is the Greek god of war. Antares means "like Ares" or "rival of Ares" (Allen 2013). Mars is the Roman god of war. Both objects are red and of similar brightness. Since the ecliptic passes through Scorpius, Mars occasionally passes very close to Antares, where they fight for dominance. We do not know the reason for the misidentification, but these could be contributing factors.

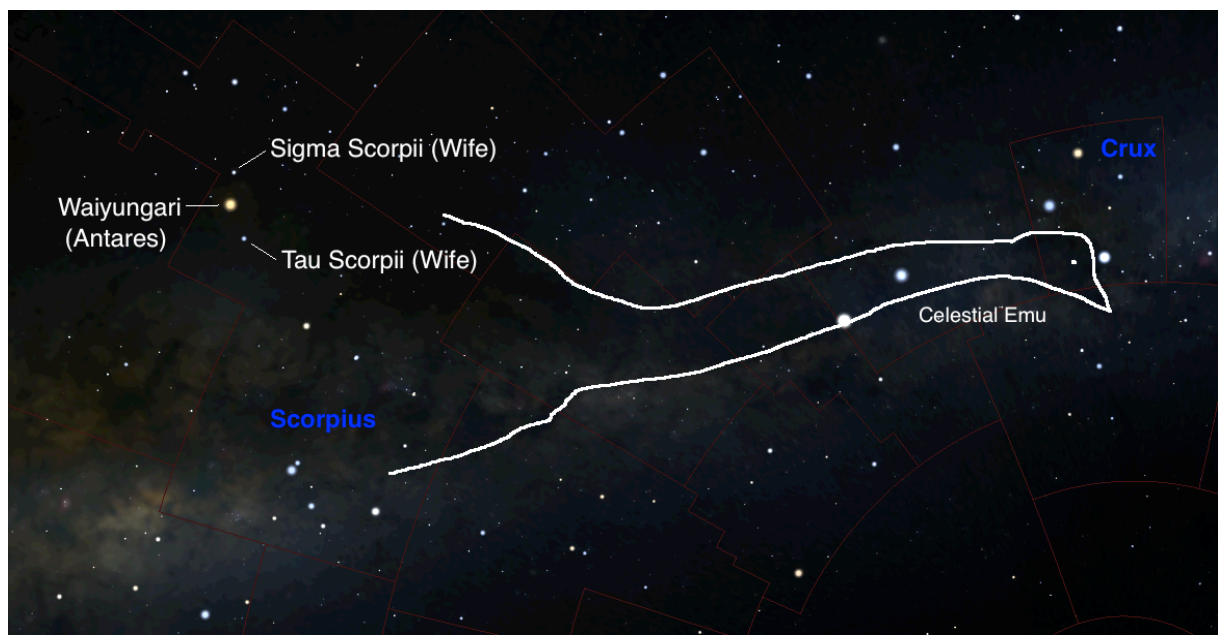

*Figure 5: Waiyungari, the unfaithful wives, and the celestial emu between Crux and Scorpius. Image modified by the author using Stellarium.*

Tindale did the right thing when it came to identifying the two celestial women. Realising his limited knowledge of astronomy, he sought out the advice of the country's top astronomer and an experienced archaeoastronomer. Since Tindale associated Waiyungari with Mars, it is understandable that the two astronomers would associate the wives with planets. A fixed star would not make sense, since the planets themselves wander along the ecliptic. Tindale (1983) stated that he did not quite understand why Dodwell or Del Chamberlain made their claim, but that he deferred to their expertise. Perhaps their opinions would have been different if Tindale told them Waiyungari was Antares.

**The Variable Nature of Nyeeruna, Kambugudha, and Waiyungari**

To be visible to the unaided eye, brightness variations in stars need to be greater than approximately 0.1 magnitudes, which is the normally accepted limit (North and James 2014). Because the apparent magnitude scale is logarithmic, the difference from one magnitude to the next represents a ~2.512 change in brightness. People with keen eyesight in ideal conditions could potentially see variations a bit fainter than this, but not substantially less. Also, Bailey and Howarth (1979, 1980) derived a correction to convert from what telescopes measure to what human eyes actually see. The corrections are as follows for the star's mean brightness: Betelgeuse (+0.80 magnitudes), Aldebaran (+1.09), and Antares (+1.35). This correction does not significantly alter the brightness changes visible to the naked eye or the results of this paper.

The Nyeeruna tradition describes the role of the stars Betelgeuse and Aldebaran in the narrative. It clearly states that Betelgeuse – the right club–wielding hand of Nyeeruna – brightens with fire magic/lust, then fades again over time. The narrative suggests the left foot of Kambugudha (Aldebaran) also brightens and fades over time, but less–so and not as frequently as Betelgeuse. Betelgeuse is a semi–regular, M–class red supergiant that varies from magnitude 0.0 to +1.3 ($\Delta$Vmag = 1.3; Figure 6–top), with a majority of its time spent at magnitude +0.50 (Samus *et al.* 2017). Variations in brightness occur with two primary periods of 388 ± 30 days and 2050 ± 460 days (Kiss *et al.* 2006). Previously published studies calculated secondary periods ranging from 1478 to 2200 days (Goldberg 1984, Wood *et al.* 2004). Aldebaran (Alpha Tauri) is a small–amplitude, slow irregular K–class orange–giant, ranging from magnitude +0.75 to +0.95 ($\Delta$Vmag = 0.2; Figure 6–middle), spending most of its time at magnitude +0.86 (Samus *et al.* 2017).

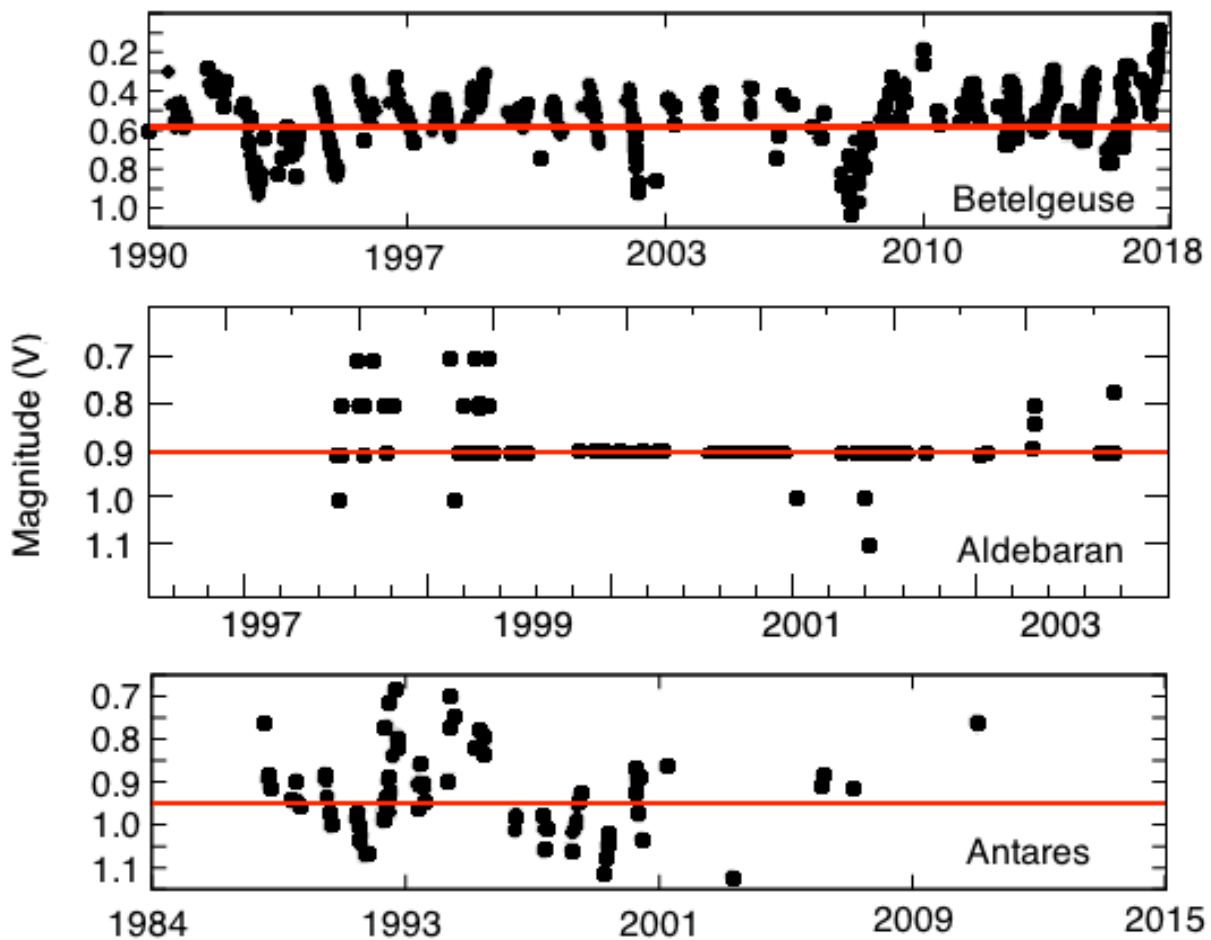

*Figure 6:* *The light curves of Betelgeuse (top), Aldebaran (middle), and Antares (bottom). Aldebaran data from VSNET observations (http://www.kusastro.kyoto–u.ac.jp/vsnet/LClast/index/TAUalpha). Betelgeuse and Antares data taken from the American Association of Variable Star Observers (https://www.aavso.org/lcg).*

The variable nature of Betelgeuse and Aldebaran is consistent with the description of Nyeeruna and Kambugudha in terms of both variable amplitude and periodicity. Described in terms of both lust and fire, the increasing passion of Nyeeruna is reflected in the brightness changes of Betelgeuse (Fredrick 2008). Both Betelgeuse and Aldebaran are described as brightening as they fill with 'fire', in association with lust and magic. Nyeeruna's right hand fades as he is humiliated, and Kambugudha's fire magic dissipates after she kicks dust into Nyeeruna's face. The periodic nature of these changes is evident in the oral tradition. Nyeeruna's 'fire lust' returns quickly, not allowing Kambugudha time to prepare. She is forced to call to the father dingo to handle Nyeeruna, after which Nyeeruna's fire lust/magic dissipates again and the brightness of his hand dims.

The Waiyungari tradition describes the Narambi becoming the star Antares (previously misidentified as Mars) and the two women become the stars Sigma and Tau Scorpii after escaping into the Milky Way to avoid punishment for breaking sacred Law. Antares is a slow irregular M–class red supergiant, ranging in brightness from magnitude +0.6 to +1.6 ($\Delta$Vmag = 1.0; Figure 6–bottom) every 1650 ± 640 days (Kiss *et al.* 2006, Samus *et al.* 2009). The star exhibits smaller peaks in brightness on a timescale of 19 years (Pugh 2013). Rather than Waiyungari brightening every September when it is visible overhead as interpreted by Tindale and the Berndts, the star's brightness peaks every few years.

Waiyungari is described as occasionally brightening, causing an increase in the sexual desire of the people, particularly *Narambi* initiates. This serves to remind the people about obeying traditional law. The sexual act between Waiyungari and the two women is reflected in the overall increase of sexual activity during the Spring. In addition to a rapid increase in breeding by animals and the flourishing of plant life, initiation ceremonies tend to be held at this time of the year (Fuller *et al.* 2013). The brightening of Antares every few years acts as a mnemonic, reminding the people to refrain from taboo forms of sexual conduct.

**Discovery and Visibility Over Time**

Definitively demonstrating that Indigenous oral traditions describe subtle astronomical phenomena can be difficult. In the case of the two oral traditions presented in this paper, the descriptions are well supported. The respective stars brighten and fade slightly over time, an observation related to the brightest stars in the sky and known to be observable to the naked eye that was incorporated into oral tradition. The periodicity of these changes is implied but not explicitly noted. The Nyeeruna tradition alludes to knowledge of the relative periodicity of Betelgeuse and Aldebaran, suggesting that Nyeeruna (Betelgeuse) fluctuates more quickly than Kambugudha (Aldebaran), as evidenced by her need to call upon Babba the father dingo to assist her as she was unable to generate the repelling fire–magic needed in time to defend against Nyeeruna's advances.

The primary peak in Betelgeuse's brightness occurs every 1.09 ± 0.08 years, with a second peak every 5.6 ± 1.26 years. Assuming a pre–colonial Aboriginal lifespan of between 40 and 60 years, Betelgeuse would have gone through 47.2 ± 9.5 primary peaks in brightness and 8.9 ± 1.8 secondary peaks. The periods of the brightness peaks of Aldebaran are not well known, but they occur irregularly. As shown it the star's light curve it can undergo frequent fluctuations over a short time, followed by long periods of inactivity.

The variability of Aldebaran is poorly studied and any established estimate of its periodicity is lacking (Wasatonic and Guinan 1997). The star is considered an irregular variable (with no regular or semi–regular brightness changes) with changes that occur less frequently than those of Betelgeuse. Aldebaran undergoes short–term fluctuations in variability, followed by longer periods of inactivity. Photometric data from VSNET reveals sporadic peaks and dips in brightness between 1997 and 2004. Sir John Herschel noted that Betelgeuse exceed Rigel in brightness at times while at other times fell fainter than Aldebaran (Herschel 1840).

Antares peaks in brightness every 4.5 years, with a secondary peak occurring every 19 years. Although slower in periodicity than Betelgeuse, this amounts to $11.1 \pm 2.2$ primary peaks and $2.7 \pm 0.6$ minor peaks during an estimated human lifespan. Thus, the variability of these stars is not a rare occurrence. The frequency of these brightness peaks would have been detected by Aboriginal people many times over a person's life. For comparison, much rarer astronomical events are well known in Aboriginal traditions. Examples include bright comets, which are visible approximately once every ten years (Hamacher and Norris 2011c), total solar eclipses, which are seen every few hundred years from a given location (Hamacher and Norris 2011a), and crater–forming meteorite impacts, which occur every few thousand years (Hamacher and Norris 2009).

It is unclear how the observations of variable stars were made by Aboriginal people. They may have utilised a technique similar to that employed by John Herschel and many contemporary variable star observers. This technique involves comparing the relative brightness of several stars of similar magnitude, utilising 'standard candle' stars that exhibit no variability. During his four–year study, Herschel (1840) noted the relative brightness of stars such as Rigel, Procyon, Acrux, Pollux, and Regulus and compared that to the stars Betelgeuse and Aldebaran.

It is difficult to know exactly when the Aboriginal observations were first made, or when they were first incorporated into oral tradition. By examining geological events in oral tradition (such as volcanic eruptions, sea level rise, and meteorite impacts) researchers have demonstrated that oral traditions can survive for thousands of years (Nunn and Reid 2016, Hamacher and Goldsmith 2014). The theoretical framework of Kelley (2015) shows how this is accomplished. With regard to the two oral traditions described in this paper, we only know the dates the stories were first published, though they are obviously much older.

With respect to variable stars, historians of astronomy attribute the first recognised discovery of a pulsating red giant to John Herschel, who observed Betelgeuse's variability between 1836 and 1840 from Cape Town, South Africa. The first record of this Nyeeruna tradition was published in the 1930s, and the first version of the Waiyungari tradition was published in 1846. These were known to be long–standing oral traditions within the Aboriginal communities at the time.

**Symbolism of the Colour Red**

When Herschel was observing Betelgeuse's variability in the 1830s, he also noted a significant increase in the brightness of the luminous blue variable star Eta Carinae. This is an unstable supergiant star that occasionally erupts, expanding in size and shedding its outer layers. This results in the star significantly increasing in brightness. In the late 1830s, Eta Carinae underwent a major eruptive event, peaking in brightness in 1843 to become the second brightest star in the night sky. This is referred to as the *Great Eruption of Eta Carinae* (Frew 2004). The Boorong clan of the Wergaia language incorporated this event into their oral traditions (Hamacher and Frew 2010). Their knowledge was recorded by William Stanbridge, who resided near Lake Tyrell in northwestern Victoria during the Great Eruption. Stanbridge kept notes of the celestial objects his Boorong informants pointed out. He recorded *Collowgullouric War* (female crow) as a bright red star, giving details of its location, appearance, and catalogue number. Although Eta Carinae is a blue star, *The Great Eruption* (as well as previous eruptions) ejected dust and debris into the surrounding space, cooling as it moved father away from the star. This scattered and attenuated the star's bluer wavelengths of light, making it appear ruddy in colour.

The colour red is seen as especially significant in Aboriginal astronomical traditions (Hamacher 2013). Aboriginal people have multiple names for red colours with an array of meanings, commonly related to sacred concepts of power, blood, and passion (Clarke 2007). Clarke argues that celestial objects that are either red or white, and bright, are associated with the power of ancestors, possessing special significance. Studies of Aboriginal astronomical traditions show that red celestial objects or phenomenon are generally assigned negative attributes. In Lardil traditions of the Wellesley Islands in the Gulf of Carpentaria, red meteors were seen as an omen of sickness, while white or other coloured meteors were seen as a sign of good news (McNight 2005: 209). In Lardil traditions, the red colour of a total lunar eclipse is associated with blood and death (Hamacher and Norris 2011a). Across Australia, the ruddy colour of the Aurora Australis is associated with war, blood, and death (Hamacher 2013).

In both oral traditions, the red colour of the stars was associated with magic and sexual lust, and this association was negative. Nyeeruna's fire magic was related to his sexual desire for the sisters of the Pleiades. His persistent attempt to rape the sisters, signified by the brightening of Betelgeuse, is only thwarted by the fire–magic and mockery of their eldest sister. In the Waiyungari story, sexual desire led to infidelity and the breaking of traditional *Narambi* law. The increasing brightness of Antares serves as a warning to the people to control their carnal urges during the *Narambi's* period of strict abstinence and to respect marriage fidelity.

The significance and symbolism attributed to red celestial objects in many Aboriginal traditions may be a contributing factor as to why the variability of Betelgeuse, Aldebaran, and Antares was noted and incorporated into oral tradition. More noticeable variations are evident in the blue star Algol, which lies 18 degrees to the north of the Pleiades and is visible across Australia. It was observed and noted by the ancient Egyptians, so why not Aboriginal Australians? It is possible Algol star *is* described in Aboriginal oral tradition, but it has either not been identified or the knowledge was lost during the last two centuries of colonisation.

Physiological and psychological factors might contribute to explaining why the variability of red stars are described in oral tradition. In terms of physiology, cones – the colour sensitive photoreceptor cells in the retina of the eye – are more sensitive to red wavelengths of light, meaning variations in brightness are more noticeable in red stars (Isles 1970). In psychological terms, both oral traditions describe a close association between the three red stars and sexual desire. Psychologists have long explored the link between the colour red and sexual attractiveness, showing evidence of their relationship (Meier *et al.* 2012). Research shows that the colour red enhances sexual attraction between people of sexual interest, in both men and women, evoking romantic approach–related motivations (Elliot and Niesta 2008, Elliot *et al.* 2010). In the context of the two Aboriginal oral traditions, the brightening red colour of these stars symbolically reinforces this relationship, reflecting traditional laws and customs relating to sex, taboos, and marriage.

**Cross–Disciplinary Training**

It is important to address any potential 'cultural contamination' between Aboriginal knowledge and Western science. The probability that Western science influenced the description of variable stars in these oral traditions is negligible. The interpretation of the narrative as relating to variable stars was not considered by ethnographers of the day and went unnoticed in the literature until Fredrick (2008) suggested the Nyeeruna tradition describes the variability of Betelgeuse, which was explored further by Leaman and Hamacher (2014). Bates (1921) attributed the increasing 'fire magic' of Betelgeuse to the "effects of radiations from nebulae". The misidentification of Waiyungari as Mars went unnoticed by astrophysicists engaged in cultural astronomy research who wrote about the tradition (Bhathal 2011, Norris 2016).

The misidentification of Waiyungari as Mars and its subsequent repetition throughout the literature by both astronomers and anthropologists highlights the necessity for cross–disciplinary training in the social and natural sciences for anyone conducting ethnographic research on Indigenous Knowledge Systems. Mistakes and errors in identifying objects can negatively affect scholarly and community outcomes. One goal of conducting work in this discipline is to accurately document and preserve Indigenous Knowledge Systems, and develop educational curricula and pedagogies from it. With very different worldviews between Aboriginal and Western ways of thinking and knowing, this can be a difficult challenge. Misidentifications, errors, and other problems that arise from having a poor understanding of astronomy can result in feeding inaccurate or faulty information back to communities. Overcoming this requires that ethnographers working on Indigenous Knowledge Systems have a sufficient working knowledge of astronomy, ecology, geosciences, and meteorology. This highlights the need for cross–disciplinary training between the natural and social sciences.

**Conclusion**

A critical re–analysis of two oral traditions from South Australia shows that Aboriginal Australians describe the variability of the red–giant stars Betelgeuse, Aldebaran, and Antares. The amplitudes in variation of Betelgeuse and Antares are conspicuous to a keen observer and are the only noticeably variable of the first magnitude stars, with a possible exception being the K–giant star Arcturus. The amplitude of Aldebaran is small but observable.

The variability of these stars measured by astrophysicists closely fits the descriptions in the Aboriginal oral traditions. The traditions imply Aboriginal knowledge about the relative periodicities in these stars' variability. I argue that Aboriginal people observed these stars and incorporated their variability into oral tradition, and that these traditions pre–date the discovery of these stars' variable nature by European scientists in the nineteenth century. This highlights the importance of considering and examining Indigenous oral traditions around the world for descriptions of celestial phenomena that can aid both astrophysicists and social scientists in their understanding of oral tradition, cultural astronomy, and Indigenous Knowledge Systems.

**Endnotes**

1. Supernovae and novae are considered 'eruptive variables'. They are caused when the star explodes or its outer layers are violently shed into space.

2. www.stellarium.org

**Acknowledgements**

I thank Philip Clarke for putting me onto the subject, and Sam Altman, Preston Borgmeyer, Bronwyn Carlson, Brad Carter, Daniel Cotton, Robert Fuller, David Harrington, Jarita Holbrook, Marcus Hughes, Trevor Leaman, Ben McKinley, Javier Mejuto, Mel Miles, Stephen

Muecke, Bruce Pascoe, Donald Reid, Jimmy Smith, Corrinne Sullivan, Rose Tasker, Ajinkya Sudhir Umbarkar, and the anonymous referees for critical feedback.

This research made use of the Stellarium software package, ViszieR stellar database, TROVE library database, the South Australian Museum archives, the SAO/NASA Astrophysics Data System, the American Association of Variable Star Observers (AASVO) light curve generator, and the VSNET database.

I acknowledge funding from the Australian Research Council (DE140101600) and pay respect to the Aboriginal communities discussed in this paper, recognising their intellectual property and traditions.

## About the Author

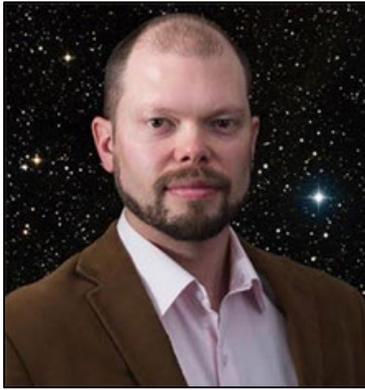

Dr Duane Hamacher is a Senior Research Fellow in the Monash University Indigenous Studies Centre and an Adjunct Fellow in the Astrophysics Group at the University of Southern Queensland. He earned degrees in physics, astronomy, and Indigenous studies and publishes extensively on Indigenous astronomical knowledge and traditions, working with elders and communities in Australia, Southeast Asia, and the Pacific. Dr Hamacher is Secretary of the International Society for Archaeoastronomy and Astronomy in Culture, Chairs the IAU C1-C4 Working Group on Intangible Heritage, serves on the IAU committee for Star Names, and is an Associate Editor of the *Journal of Astronomical History and Heritage*. He has presented at TEDx, National Geographic, the Australian Academy of Science, the BBC, ABC, and is a regular contributor to The Conversation and COSMOS Magazine.